# BRIDGING GLOBAL FRAMEWORKS: GOVERNANCE STRATEGIES BEHIND CISCO'S COMMON CONTROL FRAMEWORK (CCF) V4.0 FOR SCALABLE CLOUD COMPLIANCE

Nishant Sonkar

Cisco, Global Cloud Compliance Lead
University of Cincinnati, Ohio
San jose
Nishant19sonkar@gmail.com

## ABSTRACT

CCF v4.0 provides a standard way to ensure that Cisco's cloud products comply with the many quickly evolving requirements worldwide. To cope with increasing demands brought by ISO 27001, SOC 2, NIST, FedRAMP, EU CRA, DORA, and NIS2, CCF v4.0 introduces reliable governance by grouping controls using modules mapped across many frameworks. In this document, I discuss the governance structure controlling the framework's progress, noting how the CAB helped and the relevant steps for mapping and validating controls. Because of this, Cisco now uses the same scalable and audit-ready compliance model in all $ 10 B+ of their cloud offerings.

## 1. Introduction

The growing cybersecurity, privacy, and resilience regulations make cloud compliance more difficult. Firms that conduct business in different jurisdictions should follow and comply with standards such as ISO/IEC 27001, SOC 2, NIST 800-53, and FedRAMP, as these have some similar but different rules. Compliance with cybersecurity has become more challenging due to recent region-specific rules, for example, the EU Cyber Resilience Act (CRA), the Digital Operational Resilience Act (DORA), and the Network and Information Security Directive 2 (NIS2).

In many of these organizations, compliance efforts are spread across silos, making operations more expensive and less agile. Cisco, being at the forefront of technology and offering a broad range of cloud products, dealt with the challenge head-on. To address this issue, Cisco developed the Common Control Framework (CCF), which helps companies manage many regulations by unifying them into one set of controls.

After being established, CCF underwent several transformations that widened its area of control and improved its working methods. Version 4.0 represents a significant evolution: an organizing system that brings together past and current regulations as part of one effective compliance system.

This paper discusses the development and deployment of restructured CCF v4.0. It examines the new structural additions introduced in the current version, including the contribution of the Control Advisory Board (CAB), mapping logic across channels, and a single method for validating controls. Lastly, it examines Cisco's results across its cloud offerings and points out how CCF v4.0 can influence converging regulations for the whole industry.

## 2. Problem Statement

Expanding and complicated rules across the globe are creating more obstacles for companies to stick to the rules. The fact that many different organizations set standards is a significant problem. Although ISO/IEC 27001 stresses managing risks regarding information security, other frameworks like FedRAMP prefer specific controls to be followed in U.S. federal cloud systems. NIST 800-53 and related works also provide detailed guides and controls, which may have a distinct organization and aim than European rules like the Cyber Resilience Act (CRA), the Digital Operational Resilience Act (DORA), and the revised Network and Information Security Directive 2 (NIS2).

Every framework adds its special scope, Approach to assessment, and various enforcement tools. As a result, organizations worldwide find it hard to meet the numerous requirements put in place by different countries. Many times, security, audit, and product compliance teams have to do repeated checks on the same item, repeating assessments for every different regulatory need. If two systems are not unified, teams are more likely to face different priorities and mistakes, leading to possible breaches of regulations.

This can delay innovation progress and increase the cost and difficulty of auditing, especially for companies that work with several cloud products at once. A lack of a joined-up approach to encompass old and new rules makes it more challenging to maintain and control governance.

Upon noticing these obstacles, Cisco saw the importance of creating one compliance platform to align the rules, prevent rework, and ensure worldwide conformity. Because of this realization, developers introduced and continually developed the Cisco Common Control Framework (CCF), which led to version 4.0, which supports the organization of international governing regulations in one organized framework.

## 3. Methodology: Governance Approach

Cisco's Common Control Framework (CCF) v4.0 is based primarily on effective governance managed by the Control Advisory Board (CAB). The committee brings together people from compliance, product security, internal audit, and regulatory intelligence to oversee and manage the framework's changes. The CAB is the leading authority in controlling mapping, resolving clashes between the different frameworks, and maintaining alignment with worldwide standards as they change.

For governance, the CAB made workflows that are well-defined and structured.

- Ensuring the logical flow and proper connection between rules and regulations in each framework.
- Assenting to new controls and updates for the existing control library when facing new regulations;
- These are approaches that help countries or organizations settle differences in requirements.

I was in charge of leading the process that connected regulations to the controls required by them. I spent much time helping implement new rules, especially the EU Cyber Resilience Act (CRA), the Digital Operational Resilience Act (DORA), and NIS2. To do this, I had to look closely at their mandates, spot where Cisco's internal controls were identical, notice where they differed, and advise on changes to the existing control structure.

In addition, I ensured controls from CRA's secure development lifecycle were in line with standards, such as NIST SP 800-218, NIST SP 800-27 (Secure Application Development), and ISO/IEC 27034. As a result of their efforts, organizations in the family maintained their structure, did not overlap, and met both worldwide and regional expectations.

In addition, I contributed to adding regulations that impact healthcare, such as France's HDS and Saudi Arabia's ECC. Because of these contributions, the CCF became more effective in many different sectors and helped it remain a reliable compliance framework around the world.

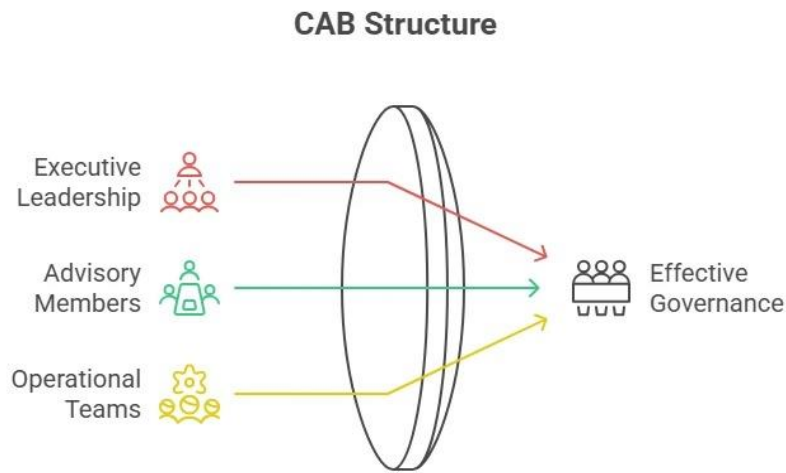

**Figure 1: The Structure of the Control Advisory Board (CAB)**

## 4. Architecture of CCF v4.0

CCF v4.0 is built from modular control families that are grouped in domains like Access Management, Incident Response, Risk Management, Cryptographic Protections, and Secure Development. Every control has a link to one or more external frameworks, creating a model with many-to-many relationships.

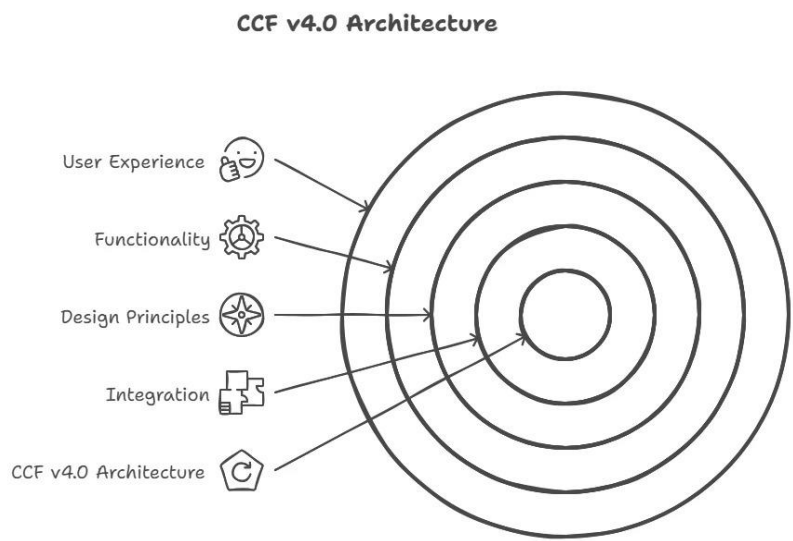

**Figure 2: Architecture of CCF v4.0**

The most significant changes in CCF v4.0 are:

- Incorporation of the controls from the EU Cyber Resilience Act (CRA) for specific products
- Putting in place ICT systems and financial risk guidelines from DORA
- Incident reporting, ensuring operations do not stop, and review of risks are required by NIS2
- Support for the French HDS and the Saudi cybersecurity guidelines has been added.

Regulations were reviewed and mapped out to see if they fit the standards already used for ISO/IEC 27001, SOC 2, and NIST CSF. This mapping method allowed both reusing and tracking existing artifacts during audits.

**Table 1. Example Mapping Snapshot**

| CCF Control ID | Control Name | Mapped Frameworks |
|---|---|---|
| CCF-IR-06 | Incident Response Testing | NIST 800-53, ISO 27001:2022, DORA, NIS2 |
| CCF-CP-04 | Cryptographic Key Rotation | FedRAMP, CRA, Saudi ECC, SOC 2 |
| CCF-SD-12 | Secure SDLC Requirements | ISO 27034, EU CRA, HDS, NIST SSDF |

## 5. Impact Assessment

Cisco's global compliance management has improved greatly since the launch of Cisco's Common Control Framework (CCF) v4.0. By combining all of its regulatory controls into one, cross-referenced system, CCF v4.0 has conveniently solved compliance needs, avoided repetitive work, and improved the company's ability to pass audits.

Key outcomes include:

Reduction in audit duplication: Cisco reduced the amount of redundant auditing in FY2024 by 35% by using a single set of artifacts for various frameworks. It was made possible by gathering all the evidence in one place and updating the way controls were assigned.

Accelerated audit response times: When control ownership and reusable documents were introduced, the time needed to prepare for external assessments dropped by 40%. Shared artifacts enabled teams to deal with auditor questions faster and prove compliance with several governance standards.

Enterprise-wide adoption and scalability: More than 300 Cisco cloud products now follow CCF v4.0, representing more than $10 billion in cloud business. Since the principles are shared by so many, the framework is efficient and valuable in various product industries and regions.

Industry recognition and external influence: Cisco's CCF is starting to shape the wider practices used in the industry. Adobe has adopted the CCF in its system of controls, and relevant groups are now holding early talks on how to align with the CCF in regulations and cloud compliance. This means that organizations in the technology sector are paying more attention to unified finance and compliance processes.

People at all levels of the organization now confirm the positive impact of CCF v4.0. The executive team has noticed the importance of the straightforward strategic Approach, traced policy, and coordinated regulations, while compliance and product specialists have stated that it has helped smooth their responses to multiple rules. Cisco has received feedback from partners showing improved trust in its compliance and governance.

## 6. Innovation & Industry Significance

The most crucial difference between Cisco's CCF v4.0 and traditional control libraries, such as HITRUST CSF and the Cloud Security Alliance's Cloud Controls Matrix (CCM), relates to its ability to anticipate changes and follow regulations as they develop. Even though most industry standards are still based on earlier codes and guidelines, CCF v4.0 already includes the latest EU regulations, such as the Cyber Resilience Act, the Digital Operational Resilience Act, and the NIS2 Directive. Because Cisco adopted these standards in the early stages, it is now ready to respond to global compliance requirements.

Several improvements in architecture and organization make the strategic value of CCF v4.0 possible.

Meta-mapping architecture: This control mapping model allows an organization to fulfill several regulatory needs using a single internal control. The architectural design enables libraries to be used efficiently and maintained less as they grow and develop.

Governance-layer abstraction: Decoupling how policies are followed from how they are interpreted allows CCF to update fast and avoid congestion in the organization. By being abstract, controls are more flexible to new rules and encourage teamwork throughout their lifecycle.

First-mover regulatory adoption: CCF v4.0 is one of the first to outline how enterprises can comply with the post-2023 cybersecurity laws in the EU. Connecting the CRA, DORA, and NIS2 standards globally in a cloud portfolio is an innovative way to achieve regulatory alignment on a mass scale.

As a result, CCF can act as a management tool and become a comparison and reference point for rivals. Because there are many differences and new regulations are challenging from country to country, international tech companies now want flexibility and trust in their audit systems. Cisco's v4.0 Compliance Framework clarifies that combining unique approaches and harmonizing multiple disciplines can help businesses remain in line with regulations.

## Conclusion

CCF v4.0 uses a strategy that helps manage cloud compliance in around the world, large organizations. With the advice of the Control Advisory Board (CAB), Cisco has handled the overlap of international standards and come up with a single, adaptable compliance model. Because of the framework, many auditors do not repeat work, controls are made consistent, and the whole structure adjusts to any new regulations.

CF can be expected to move forward by changing key aspects to address the needs of future technologies and governance.

Introducing rules and measures to ensure Ethical AI and LLM are used appropriately. With LLMs and AI systems being put into use more often, the role of fairness, transparency, and accountability in these organizations is crucial. In future, CCF is expected to include domains tailored to artificial intelligence to deal with issues of ethics and regulations in areas such as automated choices and generative tools.

Enhanced GRC Platform Integration: The focus of CCF is now on connecting more closely with Governance, Risk, and Compliance (GRC) solutions to offer real-time surveillance of compliance activities. Because of this, there can be automatic control monitoring, improved speed in handling issues, and live summaries that allow higher-ups to check on the company's regulatory situation.

Community-Led Governance and Reference Validation: In order to support industry collaboration, Cisco will allow other organizations to participate directly. This means developing community groups focused on checking the mappings and setting up reciprocal recognition plans so that everyone improves together.

Thus, the new version of CCF makes it clear that top-level leadership is crucial to keeping all global operations following the same regulatory rules. Since compliance issues are becoming more linked and flexible, the framework points out that companies should use one system for governance. Cisco wants other multinational businesses to adopt a similar model by collaborating on the creation of uniform regulations and scalable rules for the cloud.

**References**

1. Brandis, K., Dzombeta, S., Colomo-Palacios, R., & Stantchev, V. (2019). Governance, risk, and compliance in cloud scenarios. *Applied Sciences (Switzerland)*, *9*(2). https://doi.org/10.3390/app9020320
2. Yimam, D., & Fernandez, E. B. (2016). A survey of compliance issues in cloud computing. *Journal of Internet Services and Applications*, *7*(1). https://doi.org/10.1186/s13174-016-0046-8
3. Kalaiprasath, R., Elankavi, R., & Udayakumar, R. (2017). Cloud security and compliance - A semantic approach in end-to-end security. *International Journal on Smart Sensing and Intelligent Systems*, *2017*(Special issue), 482–494. https://doi.org/10.21307/ijssis-2017-265
4. Joshi, K. P., Elluri, L., & Nagar, A. (2020). An Integrated Knowledge Graph to Automate Cloud Data Compliance. *IEEE Access*, *8*, 148541–148555. https://doi.org/10.1109/ACCESS.2020.3008964
5. Apeh Jonathan Apeh, Azeez Olanipekun Hassan, Olajumoke Omotola Oyewole, Ololade Gilbert Fakeyede, Patrick Azuka Okeleke, & Olubukola Rhoda Adaramodu. (2023). GRC STRATEGIES IN MODERN CLOUD INFRASTRUCTURES: A REVIEW OF COMPLIANCE CHALLENGES. *Computer Science & IT Research Journal*, *4*(2), 111–125.https://doi.org/10.51594/csitrj.v4i2.609
6. Linkov, I., Trump, B. D., Poinsatte-Jones, K., & Florin, M. V. (2018). Governance strategies for a sustainable digital world. *Sustainability (Switzerland)*, *10*(2). https://doi.org/10.3390/su10020440
7. Mu, R., & Wang, H. (2022). A systematic literature review of open innovation in the public sector: comparing barriers and governance strategies of digital and non-digital open innovation. *Public Management Review*, *24*(4), 489–511. https://doi.org/10.1080/14719037.2020.1838787
8. Vith, S., Oberg, A., Höllerer, M. A., & Meyer, R. E. (2019). Envisioning the 'Sharing City': Governance Strategies for the Sharing Economy. *Journal of Business Ethics*, *159*(4), 1023–1046. https://doi.org/10.1007/s10551-019-04242-4
9. Driessen, P. P. J., Hegger, D. L. T., Kundzewicz, Z. W., van Rijswick, H. F. M. W., Crabbé, A., Larrue, C., … Wiering, M. (2018). Governance strategies for improving flood resilience in the face of climate change. *Water (Switzerland)*, *10*(11).https://doi.org/10.3390/w10111595
10. Collin, S. O. Y., Ponomareva, Y., Ottosson, S., & Sundberg, N. (2017). Governance strategy and costs: board compensation in Sweden. *Journal of Management and Governance*, *21*(3), 685–713. https://doi.org/10.1007/s10997-016-9359-z
11. Cheng, B., Zhu, J., & Guo, M. (2022). MultiJAF: Multi-modal joint entity alignment framework for multi-modal knowledge graph. *Neurocomputing*, *500*, 581–591. https://doi.org/10.1016/j.neucom.2022.05.058
12. Hsu, G. S., Huang, W. F., & Yap, M. H. (2020). Edge-Embedded Multi-Dropout Framework for Real-Time Face Alignment. *IEEE Access*, *8*, 6032–6044. https://doi.org/10.1109/ACCESS.2019.2960325
13. Wu, G., Guo, Y., Song, X., Guo, Z., Zhang, H., Shi, X., … Shao, X. (2019). A stacked fully convolutional network with a feature alignment framework for multi-label land-cover segmentation. *Remote Sensing*, *11*(9). https://doi.org/10.3390/rs11091051
14. Zhu, J., Huang, C., & De Meo, P. (2023). DFMKE: A dual fusion multi-modal knowledge graph embedding framework for entity alignment. *Information Fusion*, *90*, 111–119. https://doi.org/10.1016/j.inffus.2022.09.012
15. Xu, D. qin, & Li, M. ai. (2023). A dual alignment-based multi-source domain adaptation framework for motor imagery EEG classification. *Applied Intelligence*, *53*(9), 10766–10788. https://doi.org/10.1007/s10489-022-04077-z
16. Miller, C. T., Beleza, S., Pollen, A. A., Schluter, D., Kittles, R. A., Shriver, M. D., & Kingsley, D. M. M. (2007). Cis-Regulatory Changes in Kit Ligand Expression and Parallel Evolution of Pigmentation in Sticklebacks and Humans. *Cell*, *131*(6), 1179–1189.https://doi.org/10.1016/j.cell.2007.10.055

17. Klughammer, J., Romanovskaia, D., Nemc, A., Posautz, A., Seid, C. A., Schuster, L. C., … Bock, C. (2023). Comparative analysis of genome-scale, base-resolution DNA methylation profiles across 580 animal species. *Nature Communications*, *14*(1).https://doi.org/10.1038/s41467-022-34828-y
18. Liberali, P., Snijder, B., & Pelkmans, L. (2014). A hierarchical map of regulatory genetic interactions in membrane trafficking. *Cell*, *157*(6), 1473–1487. https://doi.org/10.1016/j.cell.2014.04.029
19. Liu, H., Zeng, Q., Zhou, J., Bartlett, A., Wang, B. A., Berube, P., … Ecker, J. R. (2023). Single-cell DNA methylome and 3D multi-omic atlas of the adult mouse brain. *Nature*, *624*(7991), 366–377. https://doi.org/10.1038/s41586-023-06805-y
20. Wang, Q., Xiong, H., Ai, S., Yu, X., Liu, Y., Zhang, J., & He, A. (2019). CoBATCH for High-Throughput Single-Cell Epigenomic Profiling. *Molecular Cell*, *76*(1), 206-216.e7. https://doi.org/10.1016/j.molcel.2019.07.015
21. Keromytis, A. D., & Smith, J. M. (2007). Requirements for scalable access control and security management architectures. *ACM Transactions on Internet Technology*, *7*(2).https://doi.org/10.1145/1239971.1239972
22. Singh, I., & Singh, B. (2023). Access management of IoT devices using access control mechanism and decentralized authentication: A review. *Measurement: Sensors*, *25*. https://doi.org/10.1016/j.measen.2022.100591
23. Fawcett, L., Scott-Hayward, S., Broadbent, M., Wright, A., & Race, N. (2018). Tennison: A distributed SDN framework for scalable network security. *IEEE Journal on Selected Areas in Communications*, *36*(12), 2805–2818. https://doi.org/10.1109/JSAC.2018.2871313
24. Chen, X., Feng, W., Ge, N., & Zhang, Y. (2023). Zero Trust Architecture for 6G Security. *IEEE Network*, *38*(4), 224–232. https://doi.org/10.1109/mnet.2023.3326356
25. Sutradhar, S., Karforma, S., Bose, R., Roy, S., Djebali, S., & Bhattacharyya, D. (2024). Enhancing identity and access management using Hyperledger Fabric and OAuth 2.0: A blockchain-based approach for security and scalability for the healthcare industry. *Internet of Things and Cyber-Physical Systems*, *4*, 49–67. https://doi.org/10.1016/j.iotcps.2023.07.004